\documentclass[letterpaper,11pt]{emulateapj}
\usepackage{epsfig}
\begin{document}
\title{The Star Formation History of Isolated Dwarf UGC4879}

\author{Bradley A. Jacobs \& R. Brent Tully}
\affil{Institute for Astronomy, University of Hawaii, 2680 Woodlawn Drive, Honolulu, HI 96822}

\author{Luca Rizzi}
\affil{Joint Astronomy Centre, 660 N. A'ohoku Pl, Hilo, HI 96720}

\author{Igor D. Karachentsev}
\affil{Special Astrophysical Observatory, Nizhniy Arkhyz, Karachai-Cherkessia 369167, Russia}

\author{Kristin Chiboucas}
\affil{Gemini Observatory, 670 N. A'ohoku Pl., Hilo, HI 96720}

\and

\author{Enrico V. Held}
\affil{INAF- Osservatorio Astronomico di Padova, Vicolo dell'Osservatorio, 5 - I-35128 Padova, Italy}

\begin{abstract}
Recent observations of UGC4879 with the Advanced Camera for Surveys on the Hubble Space Telescope confirm that it is a nearby isolated dwarf irregular galaxy.  We measure a distance of $1.36\pm0.03$ Mpc using the Tip of the Red Giant Branch method.  This distance puts UGC4879 beyond the radius of first turnaround of the Local Group and $\sim$700 kpc from its nearest neighbor Leo A. This isolation makes this galaxy an ideal laboratory for studying pristine star formation uncomplicated by interactions with other galaxies.  We present the star formation history of UGC4879 derived from simulated color-magnitude diagrams.
\end{abstract}

\keywords{galaxies: distances and redshifts; galaxies: dwarf; galaxies: individual: UGC4879 (VV124); galaxies: stellar content}

\section{Introduction}
The terms used to describe the environmental influence on the gas in a modest-sized galaxy found in a cluster invoke an existence that is permeated by violence.  A galaxy might lose gas through stripping, harassment, strangulation, or a combination of these or other forces.  Even the relatively quiet neighborhood of the Local Group ``is a dangerous place for dwarf galaxies'' \citep{mat98} and processes involving gas must, in many cases, be studied in the context of galaxy-galaxy interactions.  These interactions add to the complexity of a galaxy's star formation history (SFH).  Regardless of whether tidal interactions are a concern, nearly all Local Group dwarfs have been the subject of SFH studies \citep[e.g.][and references therein]{tol09} since their proximity makes it possible to resolve stellar populations inaccessible at greater distance.  The proximity and relative isolation of UGC4879 (VV124) make it an extremely attractive target for this type of study.  Here we present new observations of UGC4879 with the Advanced Camera for Surveys (ACS) on the Hubble Space Telescope and use these to construct its SFH.

\par \citet{kop08} resolved stars in UGC4879 from ground-based observations and concluded that it is on the periphery of the Local Group.  Before that, distance estimates were based on radial velocity information and tended to be erroneously high due to a value of $v_h=600\pm100$ km s$^{-1}$ assigned to it in \citet{huc83}.  As a result, UGC4879 did not receive much attention until recently.  In a followup to their 2008 paper the team in Russia extend their analysis to include spectroscopy of the unresolved stellar population and diffuse gas to confirm their measurement of $v_h=-70\pm15$ km s$^{-1}$ \citep{tik10}.  Observations of HI with the Westerbrok Synthesis Radio Telescope measure a gas velocity of $v_h=-27\pm2$ km s$^{-1}$ (Oosterloo, private communication).  We use our ACS observations of UGC4879 to extend the analysis of this galaxy.  In the next section we describe the data and reduction procedures.  From these data we derive a new distance measurement, and metallicity estimate.  Incorporating these results, we simulate the observed color-magnitude diagram (CMD) in order to reconstruct the SFH of UGC4879.  The process used to obtain the SFH is described in \S 5 and our interpretation is given in \S 6.

\section{ACS Data and Photometry}
\begin{figure}%[!ht] 
\begin{center} 
\includegraphics[width=\columnwidth]{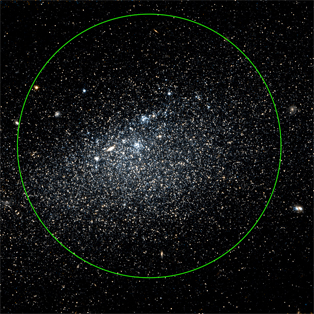}

\caption{Pseudo-color image of UGC4879 produced from $F606W$ and $F814W$ filters.  The green circle denotes the aperture within which stars were treated as members of UGC4879 for the purposes of comparison with simulated CMDs.}
\label{color}
\end{center}
\end{figure}

\begin{figure}%[!ht] 
\begin{flushleft}
\includegraphics[width=\columnwidth]{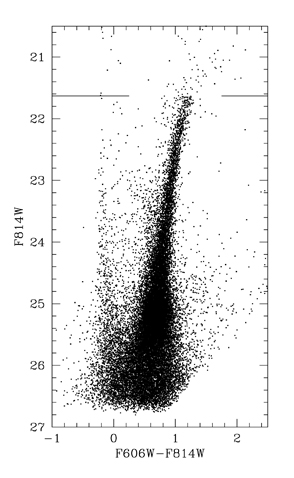}
\end{flushleft}
\begin{center}
 
\caption{CMD of UGC4879 in $F606W - F814W$ versus $F814W$ constructed from ACS observations.  The broken horizontal line at $F814W=21.63$ marks the magnitude of the TRGB.}
\label{cmd1}
\end{center}
\end{figure}

Observations of UGC4879 were carried out with ACS on February 21, 2010 in the $F606W$ and $F814W$ filters with exposures totaling 1140 s in each.  A psuedo-color image of the central portion of the ACS image created from the combined $F606W$ and $F814W$ observations is displayed in Figure \ref{color}.
We generate a stellar photometry catalog from the procedures described in \citet{jac09}.  In brief, we use the ACS module of the DOLPHOT software package\footnote{By Andrew Dolphin, http://purcell.as.arizona.edu/dolphot} to photometer the images as well as run artificial star tests to characterize the completeness and uncertainty in the measurements.   From the results of these analyses we plot the CMD in $F606W-F814W$ vs. $F814W$, see Figure \ref{cmd1}.  This diagram and all the analyses that follow are based on data within the aperture of the green circle shown in Figure \ref{color}, which is applied to reduce contamination from foreground stars.

\section{TRGB Distance}
We use a maximum-likelihood method from \citet{mak06} to measure the magnitude of the Tip of the Red Giant Branch (TRGB) in UGC4879, and find $F814W_{\rm{TRGB}}=21.63\pm0.03$.  Following the zero-point calibration of the absolute magnitude of the TRGB developed by \citet{riz07}, we calculate $M_{814}^{\rm{TRGB}}=-4.07\pm0.03$.  Using these values and accounting for foreground reddening of $E(B-V)=0.015$ from \citet{sch98} we measure a distance of $(m-M)_0=25.67\pm0.04$ or $1.36\pm0.03$ Mpc.  Further details of this distance measurement technique are available in \citet{jac09}.  Using this distance modulus with the apparent magnitude of UGC4879 of $B=13.68$ \citep{tay05} gives a (reddening corrected) luminosity of $M_B=-12.06$.

\par
\begin{figure}%[!ht] 
\begin{center} 
\includegraphics[width=\columnwidth]{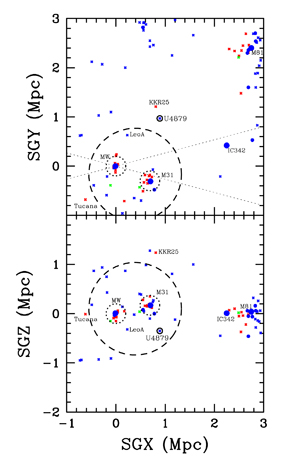}
\caption{Distribution of nearby galaxies in Supergalactic Cartesian coordinates in Mpc.  The galaxy symbols used depend on morphology and luminosity:  morphological type $T<1$ shown in red, $T\ge1$ in blue, and transitional Irr/dE in green; luminosity $M_B < -19.5$ shown as large circles, $-19.5<M_B<-17$ as small circles, and $M_B>-17$ as x's.  UGC4879, located at (0.89, 0.97, -0.35), is circled and labeled.  The large dashed circle represents the zero velocity surface of the Local Group, 940 kpc in radius, the dotted circles enclose a region 200 kpc in radius around the Milky Way and M31, and the dashed lines form a wedge at galactic latitude $|b|=15^{\circ}$.}
\label{xyz}
\end{center}
\end{figure}

 This distance is somewhat greater than the TRGB distance calculated from ground-based observations by \citet{kop08}, and puts UGC4879 just beyond the in-fall region of the Local Group rather than just within it, as they suggested.  The location of UGC4879 is plotted in Supergalactic Cartesian coordinates with reference to other nearby galaxies in Figure \ref{xyz}.  It is about equidistant from the Milky Way and M31 (1.38 Mpc), while more distant from other large nearby galaxies IC342 and M81 at 1.5 and 2.4 Mpc, respectively.  The galaxy closest to UGC4879 is Leo A, which at a separation of about 700 kpc is located on the other side of the Local Group zero-velocity surface.  This isolation indicates that its existence has been free of the type of interactions with other galaxies that could affect its SFH.

\begin{figure}%[!ht] 
\begin{center} 
\includegraphics[width=\columnwidth]{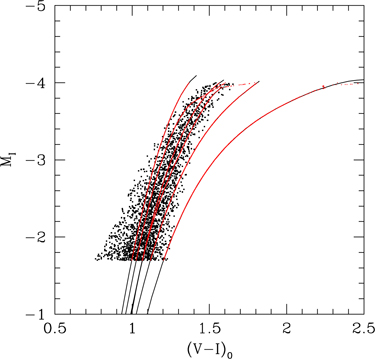}
\caption{CMD of the RGB in UGC4879 in $(V-I)_0$ versus $M_I$ selected to compare with the fiducial loci of the RGB of GGCs from \citet{dac90} which are shown as curves.  The metallicities of the GGCs used to produce these curves are:  [Fe/H]$=-2.17$, $-1.91$, $-1.58$, $-1.54$, $-1.29$, and $-0.71$.}
\label{loci}
\end{center}
\end{figure}

\section{Metallicity}
In addition to its usefulness as a distance indicator, the RGB can be used to estimate metallicity.  Our technique is based on comparing the color of the RGB stars in UGC4879 with the fiducial loci of the RGBs of the galactic globular clusters (GGCs) from \citet{dac90}.  Figure \ref{loci} shows the RGB of UGC4879 in $(V-I)_0$ vs $M_I$ plot with the Da Costa \& Armandroff loci.  After correcting for foreground reddening we calculate the mean color difference between the RGB stars and the loci of the GGCs.  We then fit a quadratic function to these resulting color differences versus the GGC metallicities.  The best fitting relation is:
\begin{equation}
[\rm{Fe}/\rm{H}]=-4.93\times \delta (V-I)^2_0-5.34\times \delta (V-I)_0-1.79
\end{equation}
with an rms of 0.05, so we take the mean metallicity of the RGB to be [Fe/H]=$-1.79\pm0.05$.  Metallicity estimates based on a sample of GGCs require that the observed population have ages comparable to the clusters, i.e. ancient stars.  In order to test this assumption with UGC4879 we generate a series of test SFHs using the methods described below, and a variety of metallicities ranging [Fe/H]=-2.0 to -0.5 and find that they consistently include large numbers of stars with ages greater than 10 Gyr.  Moreover, a star's position on the RGB is more sensitive to metallicity than age, for example an age difference of several Gyr is needed for two stars to occupy the same space on the RGB when they have a metallicity difference of order 0.1 dex \citep[e.g.][]{hel10}.

\par The GGC based value is in agreement with a metallicity estimate that depends on the $I$-band magnitude of the red clump and the distance to UGC4879 using the technique of \citet{uda00}, which uses the LMC for calibration.  Both measurements of the magnitude of stars in the RGB or in the red clump will be sensitive to old stars that are likely to have lower metallicities than younger generations of stars in the galaxy.  Hence the metallicity value quoted here should be considered an estimate of the initial metallicity rather than a global average.  In fact, as is discussed below, for the case of UGC4879 the presence of blue-loop stars on the CMD indicate recent star formation activity which is best fit with metallicities up to one dex higher than the RGB value quoted here.
\begin{figure}%[!ht] 
\begin{center} 
\includegraphics[width=\columnwidth]{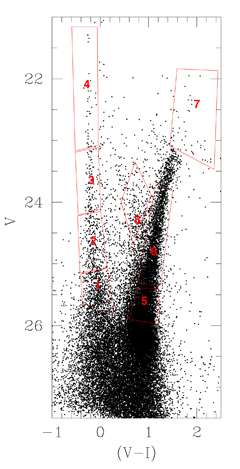}
\caption{CMD of UGC4879 in $(V-I)$ versus $V$.  The boxes enclose regions on the diagram that constrain the match between the observed and simulated CMDs.}
\label{box}
\end{center}
\end{figure}

\begin{figure}%[!ht] 
\begin{center} 
\includegraphics[width=\columnwidth]{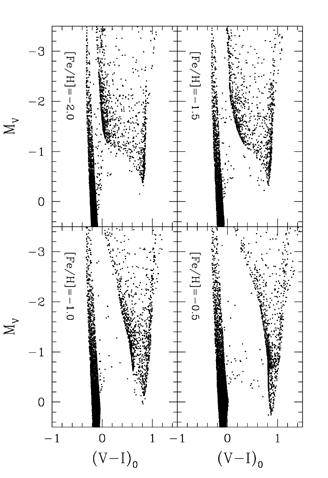}
\caption{Simulated CMDs in $(V-I)_0$ versus $M_V$ of stars with ages $<$500 Myr and fixed metallicity.  The metallicities of the stars in the panels (reading left to right) are [Fe/H]$=-2.0$, $-1.5$, $-1.0$, and $-0.5$.}
\label{blueloop}
\end{center}
\end{figure}
\section{Star Formation History}
We attempt to reconstruct the SFH of UGC4879 by simulating a CMD that matches the populations visible in the observed diagram.  The technique we use is based on the Padova evolutionary tracks \citep{ber94, gir00} and takes a number of quantities, including the initial mass function, the galaxy distance, and the reddening as input parameters.  The chemical evolution history (CEH), which is also taken as an input, is adjusted iteratively between runs to match the observed CMD as well as the metallicity information described above.  Within a single run these inputs are used to generate a set of single-age synthetic stellar populations (age bins with widths between 0.5 and 4 Gyr) using the isochrones.  Photometric errors and completeness functions are applied to the synthetic stellar populations using the results of artificial star tests produced with DOLPHOT.
We then produce a single multi-epoch CMD designed to match the observed CMD by drawing varying numbers of these degraded stars from the set of single-age populations.  The relative number of stars taken from the different age bins to match the observed CMD in turn yields the SFH.
\par The method for choosing how many simulated stars are needed from each age bin to recreate the observed CMD involves minimizing the differences in star counts between the observed and the simulated CMDs within pre-defined regions on the diagrams.  This minimization procedure is multidimensional and utilizes an Amoeba-style algorithm.   In order to increase the efficiency of this procedure and to reduce the likelihood that the resulting SFH represents a local minimum, the initial mix of stars is user-defined.  In practice, we find that the final SFH produced from a given set of input parameters converges fairly robustly even when starting from significantly divergent conditions such as a constant, continually increasing, or continuously decreasing star formation rate.  Furthermore, even a somewhat cursory look at a CMD yields information about the presence of young and old populations, so the job of the algorithm is to refine an initial mix that will usually be quite close to the final solution.
\par The errors in the SFH are measured in two ways. A first estimate is given by the minimization algorithm itself, and is based on the 99\% confidence level of the minimum. This is a lower limit to the actual errors, because it does not take into account the effect of the many random extractions that are part of the procedure.  The error contribution due to differences between subsets of synthetic stars drawn from a population that includes variations due to stellar evolution, photometric errors, and completeness limits, among others, can be quite large.  This issue becomes particularly important when matching populations of stars that exist in very low numbers on the observed CMD, such as the upper main sequence. To measure these effects, we also perform a bootstrap procedure that involves repeating the entire procedure 100 times. The final SFH is the average of the 100 repetitions, and the error is the rms of the star formation rate within each bin.  It is always the case that the bootstrap error is much larger than the minimization error.

\subsection{CMD Regions}
The predefined regions used in the minimization algorithm are selected by hand to encompass stars in various evolutionary phases.  Figure \ref{box} shows the observed CMD in $(V-I)$ vs $V$ with the regions plotted as boxes.  Boxes 1-4 mark the the main sequence stars and we attempt to include a statistically significant number of stars in each box.  Boxes 5 and 6 encapsulate samples of RGB stars.  Since the RGB includes stars of a wide range of ages, the large sample of the RGB helps to normalize the overall star formation within the galaxy.  Box 7 encloses the region where the Asymptotic Giant Branch (AGB) would be located.  The relatively low number of stars observed in this region serves to constrain the number of intermediate aged stars produced in our simulations.  Finally, in Box 8 we note the presence of blue-loop stars, which along with the high-luminosity main sequence are tracers of recent star formation.

\begin{figure}%[!ht] 
\begin{center} 
\includegraphics[width=1.05\columnwidth]{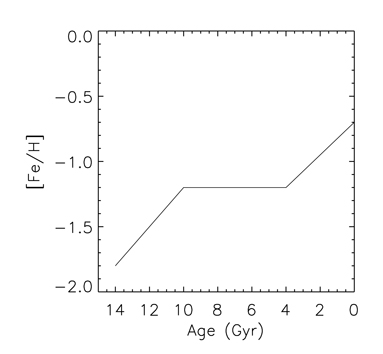}
\caption{CEH of UGC4879 used as input for constructing its SFH.}
\label{ceh}
\end{center}
\end{figure}

\subsection{Chemical Evolution History}
As an initial guess at the CEH of UGC4879 we simulate its SFH using a constant value of [Fe/H]$=-1.8$ and note which age bins show star formation activity.  From this first step we refine the CEH by requiring that eras that show significant star formation rates have a parallel increase in metallicity.  We take the RGB metallicity as measured above to be near the initial value for the galaxy.  For an endpoint metallicity at recent times we find that the location of the blue-loop stars provides a constraint.  The distribution of lower luminosity blue-loop stars between the RGB and main sequence varies as a function of metallicity.  Figure \ref{blueloop} shows a series of simulated CMDs of young ($<$500 Myr) stars where metallicity is fixed in a given panel but varies between panels.  The distribution of these lower luminosity blue-loop stars in UGC4879 (see Figure \ref{box}) appears to fall between those with [Fe/H]$=-1.0$ and [Fe/H]$=-0.5$ (lower panels of Figure \ref{blueloop}), hence we take the metallicity of young stars in UGC4879 to be [Fe/H]$\approx-0.7$. 

\begin{figure}%[!ht] 
\begin{center} 
\includegraphics[width=\columnwidth]{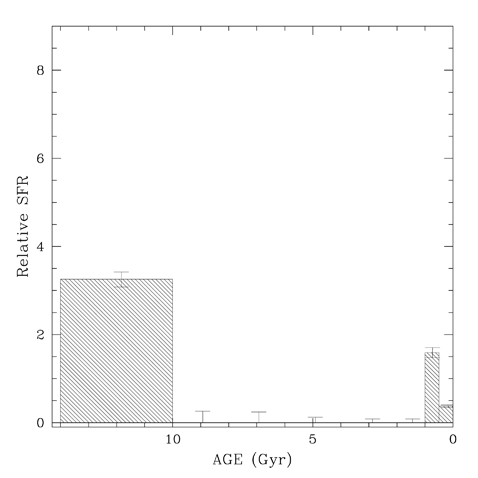}
\caption{SFH of UGC4879.  The relative star formation rate is normalized to 811 M$_{\sun}$ Myr$^{-1}$.}
\label{sfh1}
\end{center}
\end{figure}

\begin{figure*}%[!ht] 
\begin{center} 
\includegraphics[width=\columnwidth]{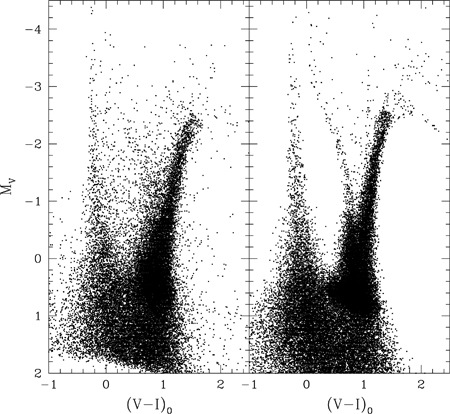}
\hspace{25 pt}
\includegraphics[width=.53\columnwidth]{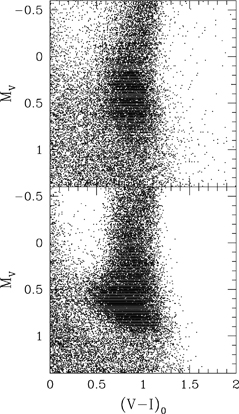}
\caption{CMDs in $(V-I)_0$ vs $M_V$ of UGC4879; the left panel consists of observed data, while the center panel consists of simulated data meant to reproduce the observed CMD based on the SFH shown in Figure \ref{sfh1}.  The right panel shows a magnified region of the CMD where the density of stars is highest.  The observed data are shown above the simulated data in this zoomed panel.}
\label{veri1}
\end{center}
\end{figure*}
\par With constraints on the endpoint metallicities of the CEH, the next area of focus becomes the nature of the increase in metallicity between these ancient and recent values.  The width of the RGB provides some information about this increase.  Although the RGB may be populated by stars of many ages, the simulations that best represent the observed CMD tend to have RGBs that are dominated by old stars.  If this is the case then the metallicity of UGC4879 must increase relatively rapidly at early times in order to account for the intrinsic width of the RGB.  Unfortunately, these data do not allow us to resolve the details of this initial star formation activity, and we note only that our simulations consistently fit the data best when we include a significant amount of star formation in the first 4 Gyr following the Big Bang.  These old stars would make up the dominant population of the RGB, and its intrinsic width is best matched with a range of metallicities of around 0.6 dex in extent.  
\par Accounting for each of these constraints, we utilize the CEH in Figure \ref{ceh} to study the SFH of UGC4879.  The best fitting SFH using this CEH is shown in Figure \ref{sfh1} and the resulting simulated CMD is shown in comparison with the observed CMD in Figure \ref{veri1}.   We work iteratively to find a CEH and SFH that are consistent with each other and together produce a simulated CMD that matches the observed.  The best matches consistently involve a CEH with (Age, [Fe/H]) at around (14 Gyr, -1.8), (10 Gyr, -1.2), and (0 Gyr, -0.7), unfortunately beyond these three estimates we lack constraints on the CEH.  One possible scenario would be that the metallicity remains flat from 10 until 1 Gyr ago, followed by a rapid increase in the last Gyr.  At the opposite extreme would be a steadily increasing metallicity between 10 Gyr and today.  Neither of these CEHs produce good matches with the data, but the CEH shown in Figure \ref{ceh}, which is intermediate between these extremes yields results that fit the data well.  Beyond this, the exact details of the CEH that we assign for ages 1-10 Gyr are not crucial because our observations are not sensitive to stars of these ages.

\subsection{Spatially Resolved Star Formation}
\begin{figure}%[!ht] 
\begin{center} 
\includegraphics[width=\columnwidth]{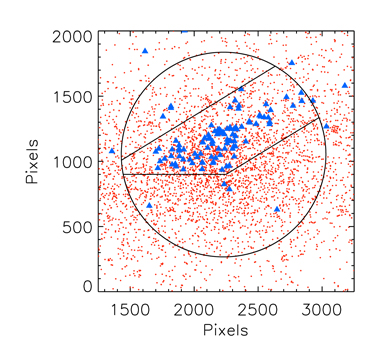}
\caption{Distribution of stars on the image in Figure \ref{color} with $I>24$.  Main sequence stars are plotted as blue triangles and RGB stars as red dots.  The circle denotes the aperture limit applied to the data to reduce foreground contamination.  The black lines denote the limits of the `inner' and `outer' regions of the galaxy for the purposes of comparing areas of active and inactive star formation.}
\label{xy}
\end{center}
\end{figure}

\begin{figure*}%[!ht] 
\begin{center} 
\includegraphics[width=\columnwidth]{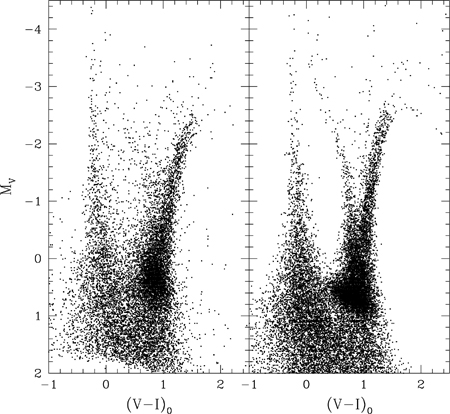}
\hspace{25 pt}
\includegraphics[width=.53\columnwidth]{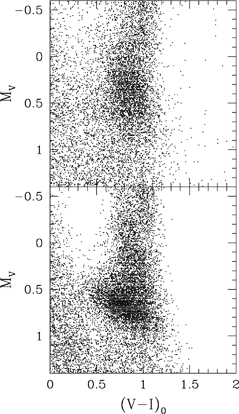}
\caption{CMDs in $(V-I)_0$ vs $M_V$ of the inner region of UGC4879 shown within the lines of Figure \ref{xy}; the left panel consists of observed data, while the center panel consists of simulated data meant to reproduce the observed CMD based on the SFH shown in Figure \ref{sfhin}.  The right panel shows a magnified region of the CMD where the density of stars is highest.  The observed data are shown above the simulated data in this zoomed panel.}
\label{inner}
\end{center}
\end{figure*}

\begin{figure*}%[!ht] 
\begin{center} 
\includegraphics[width=\columnwidth]{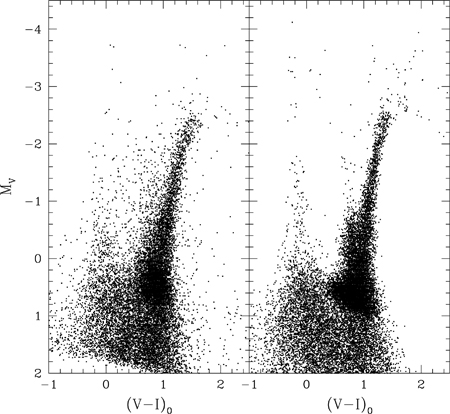}
\hspace{25 pt}
\includegraphics[width=.53\columnwidth]{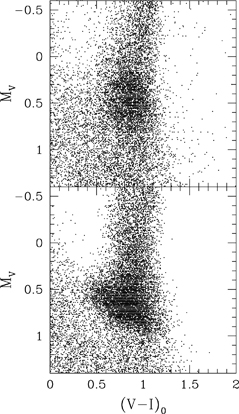}
\caption{CMDs in $(V-I)_0$ vs $M_V$ of the outer region of UGC4879 shown outside the lines of Figure \ref{xy}; the left panel consists of observed data, while the center panel consists of simulated data meant to reproduce the observed CMD based on the SFH shown in Figure \ref{sfhout}.  The right panel shows a magnified region of the CMD where the density of stars is highest.  The observed data are shown above the simulated data in this zoomed panel.}
\label{outer}
\end{center}
\end{figure*}

While we are interested in the SFH of UGC4879 as a whole, it shows signs of spatial variation in recent star formation.  Inspection of the distribution of stars in Figure \ref{color} shows that bright main sequence stars (bluish in this psuedo-color image) are highly localized.  This distribution is shown in Figure \ref{xy} where the locations of upper main sequence stars ($I<24$) are shown as blue triangles and the locations of upper RGB stars are shown as red dots.  The vast majority of the young stars, as denoted by the blue triangles, are located in a region enclosed by a set of line segments in the plot.  The areas of the regions within and outside these lines are 1603, and 3424 arcsec$^2$ respectively, which at a distance of 1.36 Mpc correspond to physical areas of $7.0\times10^4$, and $1.5\times10^5$ pc$^2$. The left panels in Figures \ref{inner} \& \ref{outer} show the CMDs of stars in the inner and outer regions.  Blue-loop and upper main sequence stars are nearly lacking from the CMD of the outer region, indicating that recent star formation has been weak or nonexistent here.  In contrast, the CMD of the inner region contains all the populations that are seen in UGC4879 as a whole.  The right panels of these figures display the simulated CMDs generated in the process of developing separate SFHs for these regions.  These SFH diagrams for the inner and outer regions are presented in Figures \ref{sfhin} \& \ref{sfhout}.  The star formation rates presented in these diagrams, as well as those measured for the entire galaxy (Figure \ref{sfh1}) are listed in Table \ref{sfrtab}.

%\vspace{3in}

\section{Discussion}
The observed CMD of UGC4879 is characterized by two distinct populations, old stars that populate the RGB with a small representation in the AGB, and young stars that make up the main sequence, blue loop, as well as contribute to the RGB, red clump and AGB.  Just how old the older stars are remains somewhat uncertain, because we do not observe the main sequence turnoff, hence cannot claim to detect an unequivocally ancient population.  The best matching SFHs consistently show significant star formation in the first 4 Gyr after the Big Bang and then a relative lull for 9 Gyr until 1 Gyr ago after which star formation is measurable again.  
\subsection{Stellar Ages $>1$ Gyr}
A question is: how significant is the 9 Gyr lull in star formation?  As mentioned above, the minimization algorithm used to define the SFH in UGC4879 takes an initial input that is user-defined.  Therefore, one way to test the significance of the relative lack of star formation at intermediate ages is to insert a star formation rate at these ages and see how things change.  Of course, doing so increases the total number of simulated stars on the CMD, so an insertion of intermediate age stars must be matched by a decrease in the star formation rate at a different age, in this case that decrease would occur in the 10-14 Gyr bin. The results from the minimization algorithm are robust under this scenario.  Any insertion of stellar populations with ages between 1 and 9 Gyr gets adjusted by the algorithm to give a star formation rate at these times that is consistent with zero.  This is because the number of simulated and observed stars in the various boxes on the CMD shown in Figure \ref{box} is matched more closely when the star formation is shifted away from these ages and into the 10-14 Gyr bin.  While it is comforting that our routine consistently reaches the same minimum, this is not an entirely satisfactory justification for the lull in star formation.  How exactly does the match improve with the exclusion of intermediate aged stars?  The key distinction comes from the number of AGB stars produced in the simulated CMD.

\begin{figure}%[!ht] 
\begin{center} 
\includegraphics[width=\columnwidth]{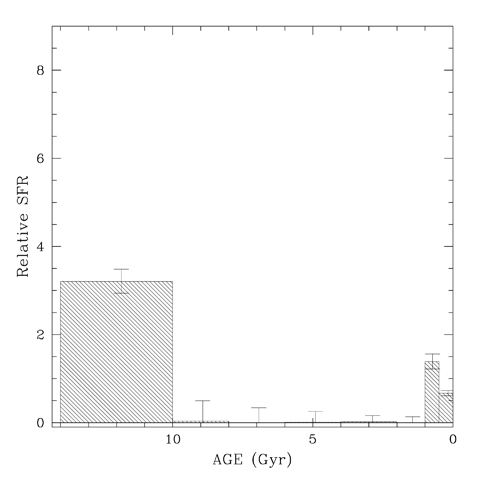}
\caption{SFH of the inner region of UGC4879 shown within the lines of Figure \ref{xy}.  The relative star formation rate is normalized to 329 M$_{\sun}$ Myr$^{-1}$.}
\label{sfhin}
\end{center}
\end{figure}

\begin{figure}%[!ht] 
\begin{center} 
\includegraphics[width=\columnwidth]{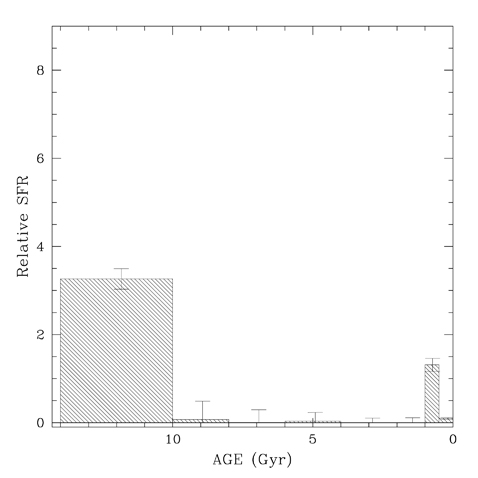}
\caption{SFH of the outer region of UGC4879 shown outside the lines of Figure \ref{xy}.  The relative star formation rate is normalized to 440 M$_{\sun}$ Myr$^{-1}$.}
\label{sfhout}
\end{center}
\end{figure}

%\clearpage

\par
\begin{figure}%[!ht] 
\begin{center} 
\includegraphics[width=\columnwidth]{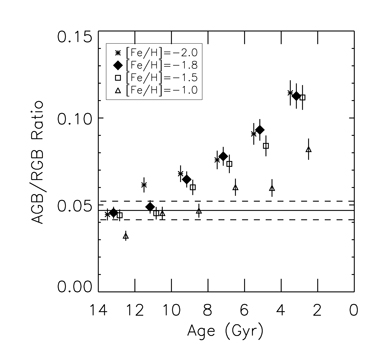}
\caption{Ratio of the number of AGB to RGB stars (within the upper 2 magnitudes of the RGB) versus age in 2 Gyr bins (the horizontal offset of the symbols within bins is for clarity only).  The asterisks,  squares, and triangles represent fixed metallicities at [Fe/H]$=-2.0$, $-1.5$, and $-1.0$, respectively, while simulated stars with the measured RGB metallicity of [Fe/H]$=-1.8$ are shown as filled diamonds.  The AGB/RGB ratio measured from observations of UGC4879 to be $0.047\pm0.005$ is plotted with horizontal lines (dashed lines indicate this uncertainty range).  The ratio is omitted in the most recent age bin because other populations provide stronger constraints on the SFH as well as add complexity to the problem of defining an RGB or AGB star on the simulated CMDs.}
\label{agbrgb}
\end{center}
\end{figure}

 At the depth of our observations there are only two stellar types distinguishable for stars with ages over 1 Gyr, those that make up the RGB and the AGB.  The relative numbers of these stars are sensitive to age, as the AGB stars evolve more quickly than those on the RGB.  In Figure \ref{agbrgb} we plot the ratio of the number of AGB to RGB stars (requiring RGB stars to be within 2 magnitudes of the TRGB to avoid confusion with other populations) versus age for simulated stars in 2 Gyr age bins.  The ratio calculated from the CMD of UGC4879 is shown as a horizontal line\footnote{For UGC4879 $I_{\rm{TRGB}}=21.64$, so we require $I<21.64$ for AGB stars and $21.64\le I <23.64$ for RGB stars included in the ratio.  In addition we apply color cuts based on visual inspection: for the AGB we require $(V-I)>1.1$ and for the RGB we use linear limits, which when written as functions have the form:  $I<-4.444\times(V-I)+27.444$ and $I>-4.444\times(V-I)+29.889$.} for an AGB to RGB ratio of $0.047\pm0.005$.  As is expected, the general trend is that the ratio increases with time and depends fairly weakly on metallicity.   The various symbols in Figure \ref{agbrgb} represent fixed metallicities at [Fe/H]$=-2.0$, $-1.5$, $-1.0$, as well as the value of $-1.8$ based on the comparison of UGC4879 with GGCs (shown as filled diamonds).  At any given time the AGB/RGB ratio is generally higher for lower metallicities, though in many cases the values are statistically indistinguishable.  We take the measured metallicity of the RGB in UGC4879 to be valid for comparing these populations.  If this is the case, then the SFH measured at ages of 10-14 Gyr is justified by the match in AGB/RGB ratios at these times.  This scenario is of course an over simplification, since for instance we see evidence for a range of metallicities in the intrinsic width of the RGB of UGC4879.  Nonetheless, the clear age evolution of the AGB/RGB ratio for fixed metallicities explains why the match to the simulations requires an early epoch of strong star formation followed by a long period of low-to-negligible star formation in UGC4879.  
\par Figure \ref{agbrgb} also shows that the AGB/RGB ratio for stars older than 10 Gyr with a metallicity of [Fe/H]$=-1.8$ is consistent with the value observed in UGC4879.  This metallicity corresponds to what was measured in the RGB of UGC4879 by comparing these stars with those in GGCs.   We present our SFH results in a combined 10-14 Gyr bin because these observations and simulations lack the resolution needed to distinguish between multiple short bursts or moderate continuous star formation at these early times.

\begin{deluxetable}{crrr}
\tablecaption{Star Formation Rates}

\tablewidth{0pt}
\tablehead{
    \colhead{Age Bins (Gyr)}&  \colhead{Full Aperture} & \colhead{Inner Region} & \colhead{Outer Region}
    }
    \startdata
0$-$0.5&303$\pm$23&220$\pm$20&44$\pm$10\\
0.5$-$1&1290$\pm$92&456$\pm$55&579$\pm$64\\
1$-$2&0$+$69&0$+$45&0$+$50\\
2$-$4&0$+$70&7$+$47&0$+$46\\
4$-$6&0$+$103&4$+$81&16$+$86\\
6$-$8&0$+$197&0$+$112&2$+$127\\
8$-$10&0$+$211&12$+$152&32$+$183\\
10$-$14&2640$\pm$139&1054$\pm$89&1438$\pm$103\\
\enddata
  \tablecomments{The absolute star formation rates in M$_{\sun}$ Myr$^{-1}$ presented in Figures \ref{sfh1}, \ref{sfhin}, \& \ref{sfhout} for the full aperture (Figure \ref{color}) and for the inner and outer regions (Figure \ref{xy}) of UGC4879.  The areas of the full, inner, and outer regions are 5027, 1603, and 3424 arcsec$^2$, respectively and correspond to physical areas of $2.2\times10^5$, $7.0\times10^4$, and $1.5\times10^5$ pc$^2$.  In age bins where the star formation rate is consistent with zero we present the upper limits and denote these as e.g. 0$+$69. \\ }
  \label{sfrtab}

\end{deluxetable} 

\par Though the AGB and RGB provide the bulk of the information on older populations observed in UGC4879, another aged population, the Horizontal Branch (HB) falls above the detection limits.  As shown in the upper panels of the zoomed CMDs in Figures \ref{veri1}, \ref{inner}, \& \ref{outer} there is a high number of stars with $M_V\sim0.5$ and $(V-I)_0\sim1.0$ observed in UGC4879.  This location corresponds to the magnitude of the HB and red clump, and most likely a mix of the two is contributing to the over-density in points, with the red clump stars extending to slightly brighter magnitudes.  The lower panels of these zoomed CMDs show the distribution of the simulated stars that make up the HB and red clump.  The simulated HB stars extend blue-ward to about $(V-I)_0\sim0.4$, while the corresponding stars in the observed CMD appear to have a more limited color range.  We have not been able to find a completely satisfactory match between the observed and simulated stars at these magnitudes.  We expect that the lack of a good match here is due to photometric errors and completeness starting to play a role in the detections, as well as that the SFH codes are unable to provide a complete picture of the HB, which represents an extremely complex stage of stellar evolution.  Despite some clear differences, the CMDs of the observed and simulated stars at these magnitudes are {\it not} wildly discrepant, and thus, while we are unable to glean additional information about the SFH here, the interpretations of other populations hold.

\subsection{Stellar Ages $<1$ Gyr}
\par The situation is clearer at recent times.  The bright main sequence and blue-loop stars observed in UGC4879 formed within the last 500 Myr.  The simulations also favor star formation between 500 and 1000 Myr ago, which apparently serves to populate the red clump.  The most recent star formation activity is confined to the central region of the galaxy as demonstrated by the difference in star formation rates in the 0-0.5 Gyr bin of Figures \ref{sfhin} \& \ref{sfhout}.  Recent studies of the SFH of nearby dwarfs have analyzed recent star formation in much finer resolution than 500 Myr \citep[e.g.][]{mcq10}, but we find that the coarser 500 Myr bins are sufficient for our purposes of localizing the area of active star formation in UGC4879.  We note also, that UGC4879 has been detected in H$\alpha$, which yields an current star formation rate of $40\pm20$ M$_{\sun}$ Myr$^{-1}$ \citep{jam04}. HI observations of UGC4879, infer a gas mass of $1.07\times10^6$ $\rm{M}_{\sun}$ (Oosterloo, private communication) and the peak of the HI emission overlaps with this region of active star formation.

\subsection{UGC4879 In Context}  
\par As would be expected for an isolated galaxy UGC4879 has not been stripped of its gas through interactions.  Its gravitational potential well is also apparently deep enough to prevent losses due to galactic winds.  This is not a forgone conclusion, as evidenced by KKR25 (LEDA 2801026), which is another fairly isolated galaxy at a distance of $1.91\pm0.06$ Mpc (see Figure \ref{xyz}) with a $B$-band absolute magnitude of $M_B=-10.0$ \citep{kar04}, but has a CMD\footnote{\raggedright The CMD of KKR25 is available through the Extragalactic Distance Database \citep{jac09}, see: http://edd.ifa.hawaii.edu/LV/KKR25/2801026info.html} that shows no signs of recent star formation.  \citet{beg05} do not detect KKR25 in HI and list a 5$\sigma$ upper limit mass of $<$$8\times10^4$ $\rm{M}_{\sun}$.  This mass limit is consistent with other properties that point toward classifying KKR25 as a dwarf spheroidal galaxy \citep{gre03}.  So it could be that it lost its gas through galactic winds, which has been suggested as the key mechanism in other isolated dwarf spheroidals \citep[e.g.][]{you07}.  Another example of a nearby isolated dwarf without recent star formation is Tucana\footnote{\raggedright For CMD of Tucana see: http://edd.ifa.hawaii.edu/LV/Tucana/69519info.html}, which is located at a distance of 920 kpc from the Milky Way, has a luminosity $M_B=-9.25$ \citep{kar04}, and no detected HI \citep{oos96}.  \citet{ber08} observe ancient ($>10$ Gyr) populations in Tucana and find no evidence for star formation in the last $\sim$8 Gyr.  It and KKR25 have SFHs that resemble that of UGC4879, except that they have no recent episodes of star formation.  Both KKR25 and Tucana are less luminous than UGC4879, but some of this difference is due to their lack of young massive stars.  Neither Tucana, KKR25 nor UGC4879 have close neighbors so the differences between their SFHs should be independent of environment.  UGC4879 would resemble the other two except, presumably, a higher mass has allowed UGC4879 to retain its gas and continue forming stars.

\par Unlike in Tucana and KKR25, HI is securely detected in UGC4879 and the contours extend well beyond the optical extent of the galaxy, possibly indicating that some gas outflow may have occurred.  \citet{tik10} also express the possibility that the HI gas is a `primodial cloud' interacting with UGC4879.  The abrupt edge to the region of star formation running from eight o'clock to two o'clock in Figure \ref{color} is suggestive of a galaxy scale compressive event.  If this is the case then this event would likely be the cause of the recent star formation.  However, if the relatively high metal content estimated in these young stars is accurate then it is doubtful that primordial gas could be the trigger for their formation.  Nevertheless, a galaxy as isolated as UGC4879 could accrete gas during its lifetime. It is possible that much of the gas involved in the current star formation could be native to UGC4879 and enriched and the compression from an external cloud might act as a trigger for the collapse of this gas.  The exact nature of the large scale gas dynamics and its affect on star formation in UGC4879 remains to be clarified.  While a lack of galaxy-galaxy interactions frees UGC4879 from some of the manifold factors that can influence SFHs, the galaxy itself contributes ample complexity to the problem.

\section{Acknowledgments}
This work was supported through HST program: GO-11584, P.I.: Kristin Chiboucas.  IDK was partially supported by the RFBR grant 10-02-00123.

\end{document}